\documentclass[12pt]{article}

\usepackage[body={17.5cm, 22.5cm},right=2cm]{geometry}

\usepackage{float}

\usepackage{booktabs}
\usepackage{caption}

\usepackage{color}
\usepackage{graphicx}
\usepackage{epsf}
\usepackage{graphicx,epsfig}
\usepackage{amsmath}
\usepackage{enumitem}

\usepackage{multirow}

\usepackage{amsmath, amsthm, amssymb}

\usepackage{epsfig}
\usepackage{cite}
\usepackage{color,colordvi}

\usepackage{hyperref}

\def\K{K{\"a}hler}

\newcounter{hran}

\allowdisplaybreaks

\makeatletter
\renewcommand\section{\@startsection {section}{1}{\z@}%
                               {-3.5ex \@plus -1ex \@minus -.2ex}%
                               {2.3ex \@plus.2ex}%
                               {\normalfont\large\bfseries}}
\makeatother

\begin{document}
\thispagestyle{empty}
\begin{flushright}
SU/ITP-14/18
\end{flushright}

\vspace{0.5cm}

\def\thefootnote{\arabic{footnote}}
\setcounter{footnote}{0}
\numberwithin{equation}{section}

\def\s{\sigma}
\def\nn{\nonumber}
\def\p{\partial}
\def\ls{\left[}
\def\rs{\right]}
\def\lc{\left\{}
\def\rc{\right\}}

\def\be{\begin{equation}}
\def\ee{\end{equation}}
\def\bea{\begin{eqnarray}}
\def\eea{\end{eqnarray}}

\newcommand{\e}{\epsilon}
\renewcommand{\l}{\lambda}
\newcommand{\lp}{\left(}
\newcommand{\rp}{\right)}
\newcommand{\bZ}{\mathbb{Z}}
\newcommand{\todo}[1]{{\bf\emph{\small {#1}}}\marginpar{$\Longleftarrow$}}

\newcommand{\bi}{\begin{itemize}}
\newcommand{\ei}{\end{itemize}}
\renewcommand{\th}{\theta}
\newcommand{\bth}{\overline{\theta}}

\newcommand{\rf}[1]{(\ref{#1})}

\def\draftnote#1{{\color{red} #1}}
\def\bldraft#1{{\color{blue} #1}}

\parskip 4pt

\begin{center}

\hskip 1cm

\vskip 1cm

{\LARGE \bf
Analytic Classes of  Metastable de Sitter Vacua
 }
\\[1.5cm]
\vskip 1.5cm
{\normalsize  \bf  Renata Kallosh,  Andrei Linde,   Bert Vercnocke and Timm Wrase
  }
\\[1.1cm]

\vspace{.1cm}
{\small {  \it Stanford Institute for Theoretical Physics, and Department of Physics, Stanford University,\\ 382 Via Pueblo Mall, Stanford, CA 94305-4060, U.S.A.}}
\\

\end{center}

\vspace{1.2cm}

\begin{center}
{\small  \noindent \textbf{Abstract}} \\[0.5cm]

\end{center}
\noindent
{\small
In this paper, we give a systematic procedure for building locally stable dS vacua  in $\mathcal{N}=1$ supergravity models motivated by string theory. We assume that one of the superfields has a \K\, potential of no-scale type and impose a hierarchy of supersymmetry breaking conditions.  In the no-scale modulus direction the supersymmetry breaking is not small,  in all other directions it is of order $\epsilon$. We establish the existence of an abundance of vacua for large regions in the parameter space spanned by $\epsilon$ and the cosmological constant. These regions exist regardless of the details of the other moduli, provided the superpotential can be tuned such that the off-diagonal blocks of the mass matrix are parametrically small. We test and support this general dS landscape construction by explicit analytic solutions for the STU model.  The Minkowski limits of these dS vacua either break supersymmetry or have flat directions in agreement with a no-go theorem that we prove, stating that a supersymmetric Minkowski vacuum without flat directions cannot be continuously deformed into a non-supersymmetric vacuum. We also describe a method for finding a broad class of stable supersymmetric Minkowski vacua that can be F-term uplifted to dS vacua and which have an easily controllable SUSY breaking scale.

\def\thefootnote{\arabic{footnote}}
\setcounter{footnote}{0}

\baselineskip= 18pt

\newpage

\tableofcontents

\newpage

\section{Introduction}
After the discovery of the cosmological constant (or dark energy) it became clear that one needs to understand the origin of de Sitter space from a fundamental theory like string theory and its corresponding effective four-dimensional space-time description. This means that one needs to understand locally stable dS vacua.
If found, a large number of such vacua including solutions with very small values of the potential at the minimum,  would allow to confront the data $\Lambda \sim 10^{-120}$ with the string theory landscape concept.
Many early attempts to do so led to various mechanisms of producing such explicit solutions. In particular, in the KKLT models \cite{Kachru:2003aw}, the volume modulus $T$, which has a no-scale \K\, potential,  is at a first stage stabilized in a supersymmetric AdS vacuum.
Then, as a second step, the minimum is uplifted to a dS vacuum. The actual value of $\Lambda$, the difference between the uplifting energy and the value of the AdS minimum can take many different values, including very small ones, which allows for an anthropic explanation of the observed value of $\Lambda$ in the context of the string theory landscape.

At present the most recent observational value for the equation of state parameter $w$ is given by the SDSS-II and SNLS supernova samples in combination with the CMB data in \cite{Betoule:2014frx}
\be
w = -1.027\pm 0.057\,.
\ee
This makes an even stronger case for the $\Lambda$CDM cosmological models where the current acceleration of the universe is due to a cosmological constant with $w=-1$. Understanding the theoretical origin of de Sitter vacua with positive cosmological constant remains a high priority.

During the last decade many interesting new constructions of dS vacua in string theory and supergravity were proposed. The ones relevant to our work are models where a single-step mechanism that stabilizes moduli in a de Sitter vacuum using only F-terms is used. In supergravity one looks for non-supersymmetric solutions where neither $W$ nor $DW$ vanish and at the minimum of the potential $V=e^K (|DW|^2 - 3|W|^2) = \Lambda  > 0$.
Early results in this direction were presented in \cite{Saltman:2004sn}. In \cite{Covi:2008ea,Covi:2008zu} a general strategy of finding dS vacua, that is similar to our own, was proposed and some explicit numerical examples of stable dS vacua in the two-moduli case were discovered.  In the simplest supergravity models that might arise in the presence of so-called non-geometric fluxes \cite{Shelton:2005cf}, numerical dS vacua where found in \cite{deCarlos:2009fq, Danielsson:2012by} and more recently in a more complicated model in \cite{Blaback:2013ht}. Other interesting ways to construct de Sitter vacua using a single step mechanism and only F-terms in string theory motivated $\mathcal{N}=1$ supergravity models have been suggested over the years, for example in \cite{Lebedev:2006qq}.
The single-step mechanism that stabilizes all moduli simultaneously in a de Sitter vacuum was also used in the recent papers \cite{Blaback:2013qza, Danielsson:2013rza}.  In particular in \cite{Blaback:2013qza} the authors used a genetic algorithm to find explicit numerical dS solutions with various properties in a type IIB string compactification that gives rise to the STU supergravity model, with three complex scalars. A similar approach in the IIA duality frame was developed in \cite{Danielsson:2013rza}. These most recent developments stimulated our work.

In this paper we describe a systematic procedure to find in string theory/supergravity families of {\it analytic, fully locally stable dS vacua} with flexible values of the cosmological constant $\Lambda$.\footnote{Our method actually leads to non-supersymmetric families for which $\Lambda$ can take positive and negative values. Interestingly, as we explain below, the range for negative values of $\Lambda$ seems much larger than the range for positive values.} We achieve this goal in the single-step procedure which by construction leads to metastable dS vacua and avoids a first stage with a supersymmetric AdS vacuum. Our de Sitter vacua, as in the KKLT case, are associated with compactifications of 10d supergravity, which also has a Dine-Seiberg solution \cite{Dine:1985he} describing 10d Minkowski space. This de-compactification limit has a vanishing vacuum energy, which means that our locally stable de Sitter vacua are globally metastable.

There are new features in our construction which have not been employed in earlier work. In particular we assume that one field has a no-scale K\"ahler potential and then do the following:
\begin{enumerate}[leftmargin=1cm,label={\arabic*)}]
\item We  introduce 2 small parameters:  $\Lambda$ is the cosmological constant and  $\epsilon$ is the scale of the supersymmetry breaking in the directions orthogonal to the no-scale field. \footnote{While these two parameters are crucial, our method generically leads to analytic dS solutions with many more parameters.}
\item We evaluate the diagonal blocks of the mass matrix at a critical point and we find that there is a region of $\Lambda$ and $\epsilon$ parameter space where they are positive definite.
\item We enforce that the off-diagonal blocks of the mass matrix vanish or are small which leads to analytic, locally stable non-supersymmetric (A)dS vacua.
\item We test our procedure in examples, like the STU models analogous to the ones studied in \cite{Blaback:2013qza, Danielsson:2013rza}. In complete agreement with our general predictions we find many new analytic examples of locally stable dS vacua.
\end{enumerate}

Our general method leads to dS vacua for which the no-scale scalar field has a mass that is always smaller than the gravitino mass. For that reason we also discuss examples of locally stable dS vacua which involve a Polonyi-type superfield, following ideas of supplementing the KKLT \cite{Kachru:2003aw} or KL scenario \cite{Kallosh:2004yh,BlancoPillado:2005fn} with a Polonyi field \cite{Dine:2006ii,Kallosh:2006dv,Dudas:2012wi}. In this second class of dS vacua, the gravitino mass can be parametrically smaller than the masses of all the scalar fields.

Along the way we prove a general no-go theorem which says that non-supersymmetric vacua cannot be obtained from locally stable supersymmetric Minkowski vacua via a small deformation of the K\"ahler and/or superpotential. This general theorem helps to explain certain features of our construction. In particular, we find in our examples that the limits to Minkowski or AdS vacua usually lead to non-supersymmetric Minkowski and AdS vacua. In some cases a limit of our dS solutions leads to supersymmetric Minkowski vacua, however, these Minkowski vacua have flat directions in agreement with the no-go theorem.

The paper is organized as follows. In section \ref{sec:strategy} we explain our general strategy for constructing families of dS vacua. The method works whenever we have (at least) one field with a no-scale K\"ahler potential and whenever we can make the off-diagonal blocks in the mass matrix small. We apply this method in section \ref{sec:STUexamples} to a variety of examples that have previously been studied in the literature. For each of these examples we can construct analytic families of dS vacua. Section \ref{sec:nogo} provides a general argument that dS vacua cannot be obtained from locally stable supersymmetric Minkowski vacua via a small deformation. In section \ref{sec:Polonyi} we discuss a different class of dS models that allows for a small gravitino mass. Our method and our results are discussed in section \ref{sec:discussion}. Appendix \ref{app:conventions} lists our conventions, appendix \ref{sec:proof} provides further details that explain our general method and appendix \ref{app:solutions} lists an explicit analytic family of dS solutions for one particular STU-model.

\section{A systematic procedure for building dS vacua}\label{sec:strategy}

In this section we give an overview and present some details of our procedure to establish conditions for locally stable de Sitter vacua in $\mathcal{N}=1$ supergravity with a set of scalar fields $X^a$. We want to find potentials $V$ such that at the extremum $V'=0$ we have $V = \Lambda >0$ and $V''$ positive-definite for a suitable set of supersymmetry breaking parameters $F_a =D_a W$.

The potential and its derivatives at an extremum $V_a = 0$ are given by \cite{Denef:2004cf}:\footnote{We use subscripts to indicate scalar derivatives. See Appendix \ref{app:conventions} for notation and further details.}
\begin{eqnarray}\label{potential}
V&=&e^{K}\Big(F_a \bar  F^a-3|W|^{2}\Big)\ ,\\
V_a&=& (D_a D_b W )\bar F^b  - 2 F_a \bar W\label{eq:dV}\,,\\
V_{ab}&=&e^{K}\Big((D_{a}D_{b}D_{c}W)\bar F^{c}-(D_{a}D_{b}W)\bar{W}\Big)\ ,\label{DDV}\\
V_{a\bar b}&=&e^{K}\Big(-{R}_{a\bar{b}c\bar{d}}\bar{F}^{c}F^{\bar{d}}+G_{a \bar b}F_{c}
\bar{F}^{c}-{F}_{a}\bar F_{\bar b}
+(D_{a}D_{c}W)(\bar{D}_{\bar{b}}\bar{D}^{c}\bar{W})-2G_{a\bar{b}}|W|^{2}\Big)\,,\label{2d}
\end{eqnarray}
where $W$ is the super potential, $K$ the \K\ potential and supersymmetry breaking is controlled by $F_a = D_a W$.

To have a local minimum, the Hessian mass matrix $M^2$ has to be positive-definite:\footnote{We will refer to the Hessian \eqref{matrixM} as mass matrix and call its eigenvalues the masses of the scalar fields. One should not that in general the \K\ metric is not the trivial Euclidean one and the kinetic terms are not canonically normalized. We do not worry about this subtlety, because a canonical renormalization of the Hessian cannot change the signs of the eigenvalues.}
\be
M^2 =
\begin{pmatrix} V_{a \bar b}  & V_{ab}\\  V_{\bar a \bar b} &  V_{\bar a b}\end{pmatrix}\,.
\label{matrixM}
\ee
A necessary condition for $M^2$ to be positive definite is that the diagonal block $V_{a \bar b} $ is positive-definite.
For models where the components of $V_{ab}$ can be made suitably small
\be
M^2_{\rm nearly \,  block-diagonal} =\begin{pmatrix} V_{a \bar b}  & \rm{small}\\  \rm{small} &  V_{\bar a b}\end{pmatrix}\,,
\label{matrixM2}
\ee
this then also ensures positive-definiteness of the full mass matrix.

Locally stable SUSY Minkowski vacua are an illustration of this principle. They are relatively easy to find, see for example \cite{Kallosh:2004yh}.
They have a non-negative definite mass matrix of the form \rf{matrixM2},  since a supersymmetric Minkowski vacuum has $W=0$ and $F_a=0$ and hence $V_{ab}=0$ (cf. \rf{DDV}). It also follows from \rf{2d} for SUSY Minkowski vacua that $V_{a\bar b}= |D^2 W|^2_{a \bar b}$ and hence it is positive-definite. As explained in \cite{BlancoPillado:2005fn}, this means that supersymmetric Minkowski vacua are either stable, for $\det D^2W\neq 0$, or  have one or more flat directions when  $\det D^2 W= 0$, but they never have tachyons. These nice properties are generic, exact and `for free' for any supersymmetric Minkowski vacuum. For dS vacua none of these nice properties are generic as explained in \cite{Marsh:2011aa}. However, as we show in this paper, for models with a no-scale field, (meta-) stable dS vacua can be constructed in abundance.

\subsection{dS vacua from nearly-no scale models}
We focus our attention on dS minima with $V = \Lambda$ and we prove the existence of a large region of locally stable dS vacua without flat directions or tachyons.  We restrict to string theory motivated $\mathcal{N}=1$ supergravity models for which one of the superfields is a no-scale modulus that we call $T$, so that $X^a= (T, X^i)$ and
\be
K = -3 \log(T+\bar T) + K(X^i,\bar{X}^{\bar{i}})\ .\label{eq:SeparableKahler}
\ee
We impose a hierarchy of supersymmetry breaking conditions $F_a(\epsilon)$
by using the small deformation parameter $\epsilon$, which controls the amount of supersymmetry breaking in the $X^i$ directions, whereas in the $T$ direction the supersymmetry breaking is not small and approximately $F_T = K_T W$:\footnote{Note that $V=\Lambda$ imposes one real relation between $F_a(\epsilon)$, $W$ and $\Lambda$. This puts one real constraint on the complex coefficient $\nu_T$ and leads to the requirement $\nu_T = \mathcal{O}[(\epsilon/|W|)^0]$.}
\be
F_T =  K_T W\ + \nu_T\frac{\epsilon^2}{|W|^2}\, , \qquad F_i = \epsilon \, \mu_i\,.\label{FT}
\ee
Here $\mu_i = F_i/ |F_i|$ and we will see later that stability requires $\epsilon \ll |W|$.

Using the hierarchy \rf{FT}, we establish that there is a large region of $\Lambda$  and $\epsilon$ parameter space where the entire diagonal block of the mass matrix $V_{a \bar b}$  is positive definite.  As long as $\epsilon>0$ this region contains positive (and negative) values for $\Lambda$. The region of stability is present for \emph{any choice of K\"ahler geometry} of the fields $X^i$ and \emph{any choice of superpotential} as long as $W$ at the minimum is non-zero: we do not require fine-tuning of $W$ and its derivatives at the minimum. There are two conditions for $V_{a  \bar b} >0$, see Appendix \ref{sec:proof} for a detailed derivation. The first one comes from requiring the sgoldstino mass $m_{sg}^2\equiv \bar F^a F^{\bar b} V_{a \bar b}/|F|^2$ to be positive:
\be
\epsilon \ll |W|\,.
\label{cond1}\ee
The second condition makes sure that the full matrix $V_{a\bar b} $ is strictly positive-definite for $\epsilon \ll |W|$:
\be
-3 e^K |W|^2 < \Lambda  < 2e^K\epsilon^2 \min_{\{L_a\}} \left(1 - \frac{(\bar \mu^i L_i)^2}{1 +  \frac{|\bar L^a D_a D_bW|^2}{|W|^2}}\right)\,, \label{cond2}
\ee
where $\{L_a\}$ is a basis of unit vectors that are all orthogonal to the supersymmetry breaking direction: $L_a \bar F^a = 0$ and $ \bar L^a L_a= 1$. Since $L_a$ and $\mu_i$ are unit vectors the right-hand-side of equation \eqref{cond2} is positive (see appendix \ref{sec:proof} for details). We thus find positivity of $V_{a \bar b}$  in large regions for $\Lambda$ and $\epsilon$.

Unlike previous proposals in the literature, we do not require a fine-tuning of the superpotential and its derivatives $W,W_a,W_{ab}$ at this point.  The lower bound on $\Lambda$ in \eqref{cond2} is of course generic and implies together with \eqref{cond1} that generically for our dS vacua (contrary to our AdS vacua) the bound on the possible values of the cosmological constant is necessarily small compared to the gravitino mass squared:  $ \Lambda < 2e^K\epsilon^2 \ll e^K |W|^2$. \footnote{This was already observed in the numerical dS vacua that were found in particular models in \cite{Blaback:2013qza, Danielsson:2013rza}.}

The typical mass scales of the $X^i$ fields are determined by the second order derivatives $W_{ab}$ in the superpotential and are generically of order $\e^0$. However, the sgoldstino mass squared $m_{sg}^2 = \bar F^a F^{\bar b} V_{a \bar b}/|F|^2$ is constrained by the condition $V_a =0$. For the hierarchy \eqref{FT} it is always small compared to the gravitino mass: $m_{sg}^2 = \frac{2}{3}e^K(2\epsilon^2-e^{-K}\Lambda)  + \mathcal{O}(\epsilon^4/|W|^2)\ll m_{3/2}^2=e^{K} |W|^2$, see eq.\ \eqref{FFV}. We also have that the mass of the field  $T$ is of the same order $m_T^2 \sim m_{sg}^2$.

In \cite{Brustein:2004xn} it was shown that with only one scalar $T$ and no other fields, one cannot get metastable de Sitter vacua. By adding a slight supersymmetry breaking orthogonal to the no-scale direction, our method circumvents this problem.
Our approach can also be applied  to a set of no-scale fields $T^\alpha$ with $K_\alpha K^\alpha = 3$, by choosing $F_\alpha = K_\alpha W + \mathcal{O}(\epsilon^2/|W|^2)$, $F_i = \epsilon \mu_i$ and hence extends previous results of \cite{Covi:2008ea} for no-scale models.

Generically, the off-diagonal blocks $V_{ab}$ can be made arbitrarily small by choosing the third derivatives $W_{abc}$ appearing in \eqref{DDV} appropriately. Then the mass matrix is of the form \eqref{matrixM2}  and the local stability of abundant dS solutions is guaranteed by the conditions \eqref{cond1}, \eqref{cond2}.

\subsection{Implementation}
We will present examples of a string theory inspired landscape with dS vacua in STU supergravities with all moduli stabilized.
The first class of examples allows us to impose the condition $V_{ab}=0$ and illustrates our general procedure, confirming that the conditions on the parameters $\Lambda$ and $\epsilon$ guarantee the positivity of the mass matrix.  The second class of examples with a simpler superpotential does not permit us to set all $V_{ab}$ to zero and some components are non-vanishing but small. Here we also find an analytic family of locally stable dS vacua.

We will use some simple, generic \K\, potentials that often arise in string theory for the volume modulus, axion-dilaton and complex structure moduli. All dependence on the parameters $\epsilon$ and $\Lambda$ enters via the superpotential.
We can expand the superpotential in a Taylor series near the minimum
\be
W = W_0 + (X-X_0)^a W_a + \frac 12 (X-X_0)^a(X-X_0)^b W_{ab} +  \frac 1{6} (X-X_0)^a(X-X_0)^b(X-X_0)^c W_{abc} +\ldots
\label{expansion}
\ee
Higher order terms are not relevant for stability, as the mass matrix only involves up to third derivatives of the superpotential as one can see from eqs. \eqref{DDV}, \eqref{2d}. A local form of our solutions is provided by giving the first terms in the Taylor series \eqref{expansion} as functions of $\Lambda$ and $\epsilon$:
\be\label{eq:TaylorCoefficients}
W_0 (\epsilon, \Lambda)\, , \qquad W_a(\epsilon, \Lambda)\, , \qquad  W_{ab}(\epsilon, \Lambda)\, ,\qquad  W_{abc}(\epsilon, \Lambda)\,.
\ee
In particular, one can infer the coefficients $W_0,W_a$ from $V=\Lambda, D_a W = F_a (\epsilon)$, the combinations $W_{ab} \bar F^b$ from $V_a =0$ through \eqref{eq:dV} and the combinations $W_{abc} \bar F^c$ from a  suitable choice of $V_{a b} $ through \eqref{DDV}.

We consider  models where the superpotential depends on the fields $X^a$ as well as on parameters $k_I$:
\be
W= W(X^a; k_I)\,.
\label{global}
\ee
Then one can rewrite the conditions \eqref{eq:TaylorCoefficients} as conditions on the parameters $k_I$. This is particularly useful when there is only a limited amount of parameters $k_I$ such that the local form near the minimum also specifies the full form of the superpotential. Then the parameters in \eqref{global} are specified in terms of $\Lambda$ and $\epsilon$:
\be
k_I (\epsilon, \Lambda)\,.
\label{sol}
\ee
We will consider examples where this can be done.
The local form of our solution is necessary and sufficient for the  study of the stability at the local dS minimum. The global form of our solutions provides more information and allows us to study models with locally stable dS vacua at different interesting regions in their moduli space. For example issues like
metastability or the height of the barriers between the metastable minima and absolute minima, require the global form of the solution. In the example we discuss in the next section $W$ is a linear function of the $k_I$ so that it is straightforward to obtain the global form of our solutions from \eqref{eq:TaylorCoefficients}.

\section{Examples in the STU model}\label{sec:STUexamples}

In this section we study a toy model with three complex scalars named $S,T,U$ that allows us to explicitly demonstrate how to build analytic dS vacua satisfying the conditions for local stability. Recently numerical examples of dS vacua without tachyons have been obtained in the four-dimensional supergravity STU-model in \cite{Blaback:2013qza,Danielsson:2013rza} and we extend these to analytic families using our method. We will take the K\"ahler and superpotential to be
\begin{equation}\label{eq:K}
K = -\log (S + \bar S) -3 \log (T + \bar T)-3 \log (U + \bar U) \,,
\end{equation}
\be
W=W(S,T,U)\,.
\ee
This model arises for example in compactifications of type IIB string theory on $T^6/\bZ_2 \times \bZ_2$. The three moduli are then the axio-dilaton $S$, the volume modulus $T$ and the complex structure $U$. This model fits into the setting of section \ref{sec:strategy} and appendix \ref{sec:proof} with $X^i = S,U$ (or $X^i = S,T$ but we focus on the former). We further discuss the relation of these STU-models to string theory in section \ref{sec:string}, where we also mention several explicit forms of the superpotential that have appeared in the literature and give rise to analytic dS solutions when using our method. In that subsection we also point out more general classes of string compactifications to which our method is applicable.

To illustrate our general method in one concrete example, we choose to work with a particular, fairly complex form of the superpotential
\begin{equation}\label{W1}
W = P(a_I,U) - S \,P(b_I,U) + A(S,U) e^{- a T}+  B(S,U) e^{- b T}\,,
\end{equation}
with
\be\label{eq:AB}
A(S,U)= P(c_I,U) - S \,P(d_I, U)\,, \qquad B(S,U) =P(e_I,U) - S \,P(f_I, U)\,,
\ee
where we take the polynomials in \rf{W1} to be all of the form
\begin{equation}\label{Ps}
P(k_I, U) = k_0 - 3 k_1 U + 3 k_2 U^2 - k_3 U^3 \,.
\end{equation}
The terms in the superpotential not involving exponentials encode the usual Gukov-Vafa-Witten flux superpotential, the exponentials model non-perturbative corrections. We give more information on the string theory origin of such a superpotential in section \ref{sec:string}.

The rest of this section is organized as follows. We first discuss extrema of the scalar potential and then we specify to two classes of examples where we find dS minima for superpotentials that are restrictions of \eqref{W1}. Our first class of examples has the feature that we can set $V_{ab}=0$, and therefore we are guaranteed to have numerous vacua as long as the conditions for $V_{a \bar b} >0$ derived in the previous section are satisfied. The second class of solutions has only one exponential term, some of the components of $V_{ab}$ are non-zero but small and we still find a large variety of locally stable dS vacua.

\subsection{Constructing dS extrema}

For 3 moduli our method requires us to impose the condition $V=\Lambda$,  to specify 3 complex conditions for $F_a(\epsilon)$, to solve 3 complex  equations for the extremum of the potential $V_a=0$,  and to impose 6 complex conditions on $V_{ab}$. In general, this would give 1 real and 12 complex equations. However, the superpotential is a holomorphic function with real coefficients. By setting the imaginary parts of $S$, $T$ and $U$ at the minimum equal to zero and choosing $\mu_i$ and $\nu_T$ to be real we trivially solve the imaginary parts of all complex equations. Furthermore, since $W$ is linear in $S$, one can prove that $V_{SS}=0$ automatically. This leads to a total of only $12$ non-trivial real equations.

We demand that the minimum for the three fields is at $S=T=U=1$. Since we have set the imaginary parts of $S$, $T$ and $U$ to zero, we can always rescale the real parameters in the superpotential so that the minimum is at any real value and the above condition is not a restriction. We make the following choice at the minimum\footnote{Note that we choose a slightly different normalization compared to eq. \eqref{FT} for ease of presentation. Now we have taken $\epsilon$ only proportional the norm of $F_i$ (in the language of section \ref{sec:strategy} we have taken $\mu_i$ not to be a unit vector but rather $|\mu_i|^2 = \bar \mu^i \mu_i = G^{U \bar U} = 4/3$). Similarly, we have set $\nu_T =|W|^2$ in $D_T W = K_T W + \nu_T \epsilon^2/|W|^2$, compared to the general form \eqref{FT}.}
\bea
V&=& \Lambda \,  , \qquad  F_T =  K_T W + \epsilon^2\,     ,\qquad  F_U = \epsilon\, ,\qquad  F_S = 0\,.
\label{eqs1}
\eea
This fixes $W_0$ and  $W_S, W_T, W_U $ at the minimum in terms of  $\epsilon$ and
\be
\lambda\equiv \frac{e^{-K} \Lambda}{2 \epsilon^2 |\mu_i|^2} = \frac{48 \Lambda}{\epsilon^2}
\ee
(cf. \eqref{eq:lambda}) as
\begin{align}
W_0 =& \frac{1}{3}(1+\epsilon^2-2\lambda)\,,\nn\\
W_S =& \frac{1}{6} (1+\epsilon^2-2\lambda)\,,\nn\\
W_T =&\epsilon ^2\,,\nn\\
W_U =&\frac{1}{2} (1+\epsilon )^2- \lambda\,.\label{eq:W_a}
\end{align}
Next we can determine three combinations of second derivatives $W_{ab}\bar F^b$ using the three stationarity equations $\partial_a V =0$ in the form given in \rf{eq:dV}. Due to our ansatz \eqref{W1} we have $W_{SS}=0$ and the critical point equations leads to three relations between the other second derivatives of $W$. Explicitly one finds that
\begin{align}
W_{ST} =&-\frac12 \e \lp 1-\e -\frac{2(2 W_{SU}-\e(1+\e))}{1-\e^2-2\l}\rp\,,\nn\\
W_{TT} =&\frac{2 \epsilon^2 \lp 2 W_{UU} -(1+(2-\e)\e-2\l)((1+\e)^2+2\l)\rp}{(1-\epsilon^2-2\lambda)^2}\,,\nn\\
W_{TU} =& \frac{\e \lp 4 W_{UU}-1-\e(5+\e(7+3\e)+6\l)+2\l \rp}{2(1-\e^2-2\l)}\,.\label{eq:W_ab}
\end{align}
From the discussion of section \ref{sec:strategy} and appendix \ref{sec:proof}, we know that for small enough $\epsilon,\lambda$ we have $V_{a \bar b}>0$. In particular, the sufficient conditions are $\epsilon \ll |W_0| \approx 1/3$ and \eqref{eq:lambdacond} for $\lambda$, which depends on the choice of $W_{SU},W_{UU}$. Note that \eqref{eq:lambdacond} puts $\lambda <1$, hence we will consider  the leading terms in a small $\lambda$-expansion in the following.

Provided the off-diagonal block of the mass matrix, $V_{ab}$, can be made small enough, the dS vacuum with small $\epsilon,\lambda$ is stable.
When third order derivatives $W_{abc}$ can be chosen at will, one can always make $V_{ab}=0$ and we have a whole range of stable dS vacua. This is class I below, for which we give a concrete realization in terms of the STU superpotential. In class II we discuss a more restricted ansatz, which does not have enough freedom in $W_{abc}$ to make all $V_{ab}$ vanish. However, we will still find stability.

\subsection{Class I: STU dS vacua with \texorpdfstring{$V_{ab}=0$}{}}

In order to satisfy $V_{ab}=0$ we have to impose 5 conditions on the third derivatives of $W$ since  it turns out that $V_{SS} =0$ automatically. There are only 7 out of 10 third derivatives $W_{abc}$ non-zero, as $W_{SSc} = 0$ for any $X^c$ due to our ansatz \eqref{W1}. In terms of $\epsilon$ and $\lambda $ we find the constraints
\begin{align}
W_{STT}  =& \frac{\e}{2} \lp -1 +2\e + \frac{4\e(2 W_{SUU}+\e^2+\e^3-2 W_{SU}(2+\e))}{(1-\e^2-2\l)^2} + \frac{2 (2 W_{SU}-\e)}{1-\e^2-2\l}\rp \,,\nn\\
W_{STU}  =& \frac{4 W_{SU}(1-\e-2\e^2-2\l)+\e(8W_{SUU}-2+4\l)-\e^2(2+8 \l) -2\e^3-\e^4-(1-2\l)^2}{4(1-\e^2-2\l)}\,,\nn\\
W_{TTT}  =& \frac{2\e^3 (4 W_{UUU}-6W_{UU}(1+\e)+(1+\e)^3(2+(3-\e)\e))}{(1-\e^2-2\l)^3}\nn\\
&-\frac{2\e^2 \l((1-2\l)^2)+4\e+\e^2(2+8\l)+5\e^4}{(1-\e^2-2\l)^3}\,,\nn\\
W_{TTU}  =& \e \lp -1 +3\e+\frac{2\e(2W_{UUU}-\e(3+2 W_{UU}-3\e^2))}{(1-\e^2-2\l)^2} +\frac{2 W_{UU}-2\e(4+\e)}{1-\e^2-2\l}\rp\,,\nn\\
W_{TUU}  =& W_{UU}-\frac12(1+\e(4-3\e)) +\frac{\e \lp 2W_{UUU} + W_{UU}(1-\e)-\e(5-\e)(1+\e)\rp}{(1-\e^2-2\l)}+\l\,.\label{eq:W_abc}
\end{align}
All dS vacua of the STU model \eqref{W1} with $V_{ab}=0$ obey these relations between their third order derivatives. For presentational purposes, we have chosen the four unconstrained values to be $W_{SU}$, $W_{UU}$, $W_{SUU}$ and $W_{UUU}$.

An alternative description of our analytic dS solutions is not using the expansion near the minimum as in eq. \rf{expansion} but by providing the explicit expressions for the values of $a_I, b_I, c_I, d_I, e_I, f_I$ in eq. \rf{W1}. We discuss one particular randomly chosen minimal example. Since we have in total 12 real constraints on $W_0,W_a,W_{ab}, W_{abc}$, we make an ansatz with 12 flux parameters, in addition to the two exponents $a,b$:
\be
W = (a_0 -3 a_1 U + 3 a_2 U^2 - a_3 U^3)  - S(b_0 -3 b_1 U + 3 b_2 U^2 ) + (c_0 -3 c_1 U + 3 c_2 U^2  -d_0 S) e^{-aT}+ e_0 e^{-bT}\, ,
\label{simple1}\ee
where we have simply set several of the parameters in \rf{W1} to zero. Then the 12 equations \eqref{eq:W_a}, \eqref{eq:W_ab}, \eqref{eq:W_abc} determine all parameters in terms of $\epsilon,\lambda,a,b$: the local  vacuum analysis fixes the global form of the superpotential \eqref{simple1}.

The full expression is lengthy and we present it in an accompanying Mathematica notebook titled \mbox{\texttt{Solutions.nb}} that makes use of the code written in \cite{Kallosh:2004rs}. A representative can be shown here, for example
\be
e_0= \frac{2(1+a) e^b \epsilon^2 (\epsilon^2+ 3 \lambda)}{(a-b) b (1+3\epsilon +b( -1+\epsilon^2+2\l) -2\l)}\,.
\ee
It is interesting to notice that for a single exponent in $W$ which would be the case for $a=b$ this model would have no solution with $V_{ab}=0$.

Now that we know that we have dS vacua we want to show explicitly that they are locally stable in a certain region of  $\epsilon$ and $\Lambda$ parameter space. The mass-matrix is block diagonal since  $V_{ab}=0$:
\be
M^2 =
\begin{pmatrix} V_{a \bar b}  & 0\\  0&  V_{\bar a b}\end{pmatrix}\ .
\label{matrix}
\ee
Since we consider a vacuum at real values  $S=T=U=1$ the two blocks are identical and there are three doubly degenerate mass eigenvalues. We denote the three eigenvalues as $m_{S}^2,m_T^2,m_U^2$, as to leading order in $\e$ the eigenvectors align with the moduli directions. From the general discussion of section \ref{sec:strategy} and appendix \ref{sec:proof}, we know that $\epsilon \ll |W| \approx 1/3$ is a sufficient condition for stability and we will see that $\epsilon$ can actually be close to $1/3$. The condition \eqref{cond2} on $\Lambda$ or equivalently on $\lambda = 48 \Lambda/\epsilon^2$ requires a little more work.

To zeroth order in $\epsilon$ and to first order in $\l$ the eigenvalues of $V_{a \bar b}$ are
\be
m_{S}^2 = \frac{1}{4608}(1-4\lambda)  ,\quad m_{T}^2 = 0\ , \quad m_{U}^2 =  \frac{1}{1536}\lp a^2+1-4\lambda\ \frac{  a (a+1) (2 b+1)+b-1}{b-1}\rp\ .\label{FirstOrder1}
\ee
The values for $m_{T}^2$ at higher order give very lengthy expressions. We note that $V_{a \bar b}$ is to leading order in $\epsilon$ a diagonal matrix so that the necessary and sufficient conditions for positive eigenvalues can be written as $m_{S}^2>0, m_U^2>0$ and $\det V_{a \bar b}>0$:
\be
0 < \det V_{a \bar b} = \frac 1 {2^{23} \cdot 3 ^4}\epsilon ^2\left[ a^2 -\lambda\ \frac{ a^2 (13 b-1)+a (8 b+4)+b-1}{(b-1)} \right]+ \mathcal{O}(\epsilon^3,\lambda^2)\ .\label{eq:classIcond}
\ee
One finds that for $a\geq0, b>1$ the most stringent of these three conditions is \eqref{eq:classIcond}, corresponding to \eqref{cond2} for this specific model. We illustrate this with six examples in Table \ref{tab:ClassI} for $a=1,b=2$, for which the three conditions $(m_{S}^2,m_U^2,\det V_{a \bar b})>0$ become:
\be
{\Lambda} = \frac{\l \e^2}{48} <\frac {\epsilon^2}  {48 }\min \left(\frac 1{4}, \frac 1 {22}, \frac 1 {46}\right) = \frac {\epsilon^2}  {2208 }\ .\label{eq:Lambda23}
\ee

\begin{table}[ht!]
\centering
\begin{tabular}{|c|c|c|c|c|}
\hline
$\epsilon$&$\Lambda$&$m_{\text{Re(S)}}^2=m_{\text{Im(S)}}^2$ & $m_{\text{Re(T)}}^2=m_{\text{Im(T)}}^2$ & $m_{\text{Re(U)}}^2=m_{\text{Im(U)}}^2$\\
\hline
$10^{-1}$&$10^{-6}$&0.000218608&0.0000436152&0.00109265\\
\hline
$10^{-2}$&$10^{-10}$&0.000217016&$5.20265\cdot10^{-7}$&0.00130028\\
\hline
$10^{-2}$&$10^{-120}$&0.000217057&$5.20775\cdot10^{-7}$&0.00130168\\
\hline
$10^{-3}$&$0$&0.000217014&$5.20833\cdot10^{-9}$&0.00130208\\
\hline
$10^{-2}$&$2\cdot 10^{-6}$&$0.000184146$&$3.70445\cdot10^{-8}$&0.140283\\
\hline
$10^{-3}$&$10^{-6}$&1.95855&$-4.89583\cdot10^{-7}$& $4.37582\cdot10^6$\\
\hline
\end{tabular}
\caption{\small Eigenvalues for several choices of $\epsilon,\Lambda$ for $a=1,b=2$. These should be compared to the zeroth order results \eqref{FirstOrder1} $m_{\text{Re(S)}}^2=m_{\text{Im(S)}}^2 =1/4608\approx 0.000217014$, $m_{\text{Re(U)}}^2=m_{\text{Im(U)}}^2 = 1/768\approx 0.00130208$ and the leading order result $m_{\text{Re(T)}}^2=m_{\text{Im(T)}}^2 =\e^2/192 = 0.00520833 \cdot \e^2$. The first four examples all obey the condition for stability at lowest order in $\lambda,\epsilon^2$ derived in \eqref{eq:classIcond} and \eqref{eq:Lambda23}: $\Lambda < \frac {\epsilon^2}  {2208}\approx 4.53 \cdot10^{-4}\epsilon^2$. The fifth example has $\lambda = 48 \Lambda/\epsilon^2 = 0.96$, so the series expansion in $\lambda$ is not justified, however we still find stability. The last example has $\lambda = 48 \Lambda/\epsilon^2 = 48$ and the series expansion completely breaks down, leading to two very large and one negative eigenvalues. Note that the negative eigenvalue is due to the negative terms with $\Lambda$ in eq.\ \eqref{FFV} dominating the $\epsilon$-dependent terms.
\label{tab:ClassI}}
\end{table}

\subsection{Class II: STU dS vacua with some \texorpdfstring{$V_{ab}\neq0$}{}}\label{sec:SimpleSolution}
We now study a simplified version where only one exponential term for $T$ is present and where in front of this exponential term we have only  $U$-dependence. This smaller set of parameters in the superpotential is still sufficient to present examples of analytic dS vacua that are locally stable. The superpotential is
\begin{equation}\label{W1s}
W = P(a_I,U) - S P(b_I,U) + P(c_I,U)  e^{- a T}\,.
\end{equation}
Models of this type were studied numerically in \cite{Blaback:2013qza}.
Since the $P$'s are cubic polynomials in $U$ (cf. equation \eqref{Ps}), our superpotential has 12+1 real parameters: $a_I, b_I, c_I$ and $a$:
\be
W = (a_0 -3 a_1 U +3 a_2 U^2 - a_3 U^3)  - S(b_0 -3 b_1 U + 3 b_2 U^2 - b_3 U^3) + (c_0 -3 c_1 U + 3c_2 U^2 - c_3 U^3) e^{-aT}\ .
\label{simple}\ee
The parameters $b_3,c_3$ encode fourth order derivatives and do not play a role in the stability analysis and we will put them to zero. Hence the 12 conditions $V=\Lambda, D_a W = F_a(\epsilon), V_a=0,V_{ab}=0$ give an over-constrained system for the 11 remaining parameters. We want to leave the exponent $a$ as a free parameters and find two components $V_{ST}$ and $V_{TT}$ that we cannot put to zero.

We write the  conditions \rf{eqs1} in terms of the superpotential and its (up to third) derivatives. Since $S$ appears only linear in the superpotential we have $W_{SS} = W_{SSc}= 0$ for any $X^c$. Since there is only one exponent $e^{-aT}$ with an $S$-independent coefficient, we have $W_{ST} = 0$ and $W_{TT}=-a W_{T}=-a\e^2$, which together with the constraints \eqref{eq:W_ab} fixes all second derivatives of $W$ at the minimum. In our simple model we also have $W_{TTc} =-a W_{Tc}$. Therefore there are only three unconstrained third derivatives which can be used to set $V_{SU}=V_{TU}=V_{UU}=0$. We then find
\begin{align}
W_{SUU}  =& \frac{1}{4} \left(\left(1+\epsilon ^2\right) (1+\epsilon )^2-\lambda  (3-\epsilon  (4+3 \epsilon ))+2 \lambda ^2\right)\,,\nn\\
W_{TUU}  =& \frac{1}{4} \Big{(}a^2 \left(1- \epsilon ^2-2\lambda\right)^2-a (1 +\epsilon  (4+3 \epsilon )+4 \lambda) \left(1- \epsilon ^2-2\lambda\right) +2 \epsilon  \left((2+\epsilon) (1+\epsilon )^2+2\lambda  (4+\epsilon)\right)\Big{)}\,,\nn\\
W_{UUU}  =& \frac{\epsilon  (1+\epsilon )^2 (1+3 \epsilon  (1+(1-\epsilon) \epsilon))-2\lambda(1 -\epsilon  (2+\epsilon  (6-\epsilon  (6+7 \epsilon))))+4\lambda ^2 (1+\epsilon ) (2-5 \epsilon)-8\lambda ^3}{4 \epsilon}\nn\\
&+\frac{a^2 \left(1- \epsilon ^2-2\lambda\right)^3 - a \left(1- \epsilon ^2-2\lambda\right)^2 (\epsilon  (3+5 \epsilon )+6 \lambda )}{8 \epsilon }\,.
\label{class2}\end{align}
The component $V_{SS}$ is automatically zero, while the expressions for $V_{TT}$ and $V_{ST}$ are non-zero and given by
\be
V_{ST} = -\frac{1}{384} \epsilon ^2 \left(1+\epsilon ^2+4 \lambda \right)\, ,  \qquad V_{TT} = -\frac{1}{96} (1+a) \epsilon ^2 \left(\epsilon ^2+3 \lambda \right)\ .
\label{nonzero}
\ee
These are order $\epsilon^2$ and especially $V_{TT}$ will affect the stability bound on $\lambda$. We will see below that these components are still small enough to allow for a large range of stable solutions.

Our solutions can be given in the form of explicit expressions for $a_I, b_I, c_I$ as functions of $\epsilon$ and $\lambda$ and $a$, see appendix \ref{app:solutions}.

The leading contribution to the mass eigenvalues are the same as for class I with $b\to \infty$:
\be
m_{S}^2 = \frac{1}{4608}(1-4\lambda) + \mathcal{O}(\epsilon)\ ,\quad m_{T}^2 = 0 + \mathcal{O}(\epsilon^2)\ , \quad m_{U}^2 =  \frac{1}{1536}\left(a^2+1-4\lambda(1+2a +2 a^2)\right)+ \mathcal{O}(\epsilon)\ .\label{FirstOrder2}
\ee
Note that at higher order in $\epsilon,\lambda$ the degeneracies of the eigenvalues are lifted.
There are two small eigenvalues $m^2_{\text{Re(T)},\text{Im(T)}}$ which also need to be positive. At small $\lambda$ they are approximately
\be
m^2_{\text{Re(T)}} \approx m_{\text{Im(T)}}^2 =\frac 1{96} \frac{a^2}{(a^2 + 1)} \epsilon^2 + \mathcal{O}(\lambda,\epsilon^3)\,.\label{eq:massT}
\ee
Evaluating higher order terms for general values of the parameter $a$ gives an unwieldy expression.

Instead, we note that $M^2$ is positive-definite provided all its upper-left diagonal square sub-matrices are positive-definite. One finds this does not lead to a stronger condition on $\lambda$ than $m_U^2 >0$ or in terms of $\Lambda$:
\be
\Lambda = \frac{\l \e^2}{48} < \frac{\epsilon^2}{192}\frac{1 + a^2}{(1 + 2 a + 2 a^2)}\label{eq:lambdaII}\,.
\ee
Again we establish  a continuous range of locally stable de Sitter vacua for small $\Lambda,\epsilon$. We give some specific values in table \ref{tab:ClassII}.

\begin{table}[ht!]
\centering
\begin{tabular}{|c|c|c|c|c|c|c|c|}
\hline
$\epsilon$&$\Lambda$&$m_{\text{Re(S)}}^2$&$m_{\text{Im(S)}}^2$ & $m_{\text{Re(T)}}^2$&$m_{\text{Im(T)}}^2$ & $m_{\text{Re(U)}}^2$&$m_{\text{Im(U)}}^2$\\
\hline
$10^{-1}$&$10^{-6}$&0.000220&0.000228&0.0000723&0.0000797&0.00321&0.00321\\
\hline
$10^{-2}$&$10^{-10}$&0.000217&0.000217&$8.32\cdot 10^{-7}$&$8.33\cdot 10^{-7}$&0.00325&0.00325\\
\hline
$10^{-2}$&$10^{-120}$&0.000217&0.000217&$8.33\cdot 10^{-7}$&$8.33\cdot 10^{-7}$&0.00326&0.00326\\
\hline
$10^{-3}$&$0$&0.000217&0.000217&$8.33\cdot 10^{-9}$&$8.33\cdot 10^{-9}$&0.00326&0.00326\\
\hline
$10^{-2}$&$2\cdot 10^{-6}$& $0.000184$&$0.000184$&$-8.99\cdot10^{-6}$&$9.00\cdot10^{-6}$&$0.0183$&$0.0183$\\
\hline
$10^{-3}$&$10^{-6}$&1.96&1.96&$-4.99\cdot10^{-6}$ & $4.01\cdot10^{-6}$&$8.57\cdot10^5$& $8.57\cdot10^5$\\
\hline
\end{tabular}
\caption{\small
Eigenvalues for several choices of $\epsilon,\Lambda$ for $a=2$. This should be compare to the zeroth order results \eqref{FirstOrder2} $m_{\text{Re(S)}}^2 = m_{\text{Im(S)}}^2 =1/4608\approx 0.000217, m_{\text{Re(U)}}^2 = m_{\text{Im(U)}}^2 = 5/768 \approx 0.00326$ and the lowest order result \eqref{eq:massT} $m_{\text{Re(T)}}^2 =  m_{\text{Im(T)}}^2 = \epsilon/120 \approx 0.00833 \epsilon^2$. The first four examples all obey the condition for stability at lowest order in $\lambda,\epsilon^2$ derived in \eqref{eq:lambdaII}, which becomes for $a=2$: $\Lambda < \frac {5\epsilon^2}  {2496}\approx 2.00 \cdot10^{-3}\epsilon^2$. The fifth example has $\lambda = 48\Lambda/\epsilon^2 = 0.96$, so the series expansion in $\lambda$ is not justified and we find a tachyonic direction. The last example has $\lambda = 48\Lambda/\epsilon^2 = 48$ and the series expansion completely breaks down, again giving an instability.
\label{tab:ClassII}}
\end{table}

\subsection{Relation to string theory and more general models}\label{sec:string}

The four-dimensional supergravity STU-model that we have used to illustrate our general approach and whose K\"ahler and superpotential are given in \eqref{eq:K} and \eqref{W1} or \eqref{W1s}, arises in compactifications of type IIB string theory on $T^6/\bZ_2 \times \bZ_2$, if we restrict to the isotropic sector. The compactification of type IIB string theory on this toroidal orbifold leads to a four-dimensional theory that has three moduli $S, T$ and $U$ and a K\"ahler potential as given in \eqref{eq:K}. The modulus $S$ corresponds to the axio-dilaton. Its real part is given by the inverse string coupling and its imaginary part by the $C_0$-axion of type IIB string theory $S = e^{-\phi} + i C_0$. The real part of the field $T$ controls the size of the internal space and its imaginary part is given by the four dimensional axion field that arises from reducing $C_4$. Finally, the complex field $U$ controls the complex structure of the three identical $T^2$'s.

In order to generate a superpotential we can turn on fluxes that thread the six compact internal directions. We can turn on four independent $F_3$ flux quanta which correspond to the $a_I$'s in our superpotentials  \eqref{W1} or \eqref{W1s}. We can also turn on four different $H_3$ flux quanta that are in a one to one correspondence with the $b_I$'s. Lastly, we would like to generate one or more exponential terms for $T$. Such terms can either arise from gaugino condensation on D7-branes or from Euclidean D3-branes.\footnote{The D7-branes or Euclidean D3-branes wrap invariant combinations of two of the three $T^2$, while the $H_3$ flux has one leg along each of the three $T^2$'s so that there is no Freed-Witten anomaly.} In both cases the superpotential receives a contribution of the form
\be
W_{np} = A(S,U) e^{-a T}\,,
\ee
where $a =2\pi$ for an Euclidean D3-brane and $a=2\pi/N_c$ for a stack of $N_c$ D7-branes. While it is in principle possible to calculate the explicit functional form of $A(S,U)$ for the toroidal orbifold $T^6/\bZ_2 \times \bZ_2$ we refrain from doing so here (see \cite{Berg:2004ek} for a related calculation at fixed string coupling in this model). In our explicit examples we assume a certain simple dependence of the prefactor on $U$ and $S$ (cf. \eqref{simple1} and \eqref{W1s}) for concreteness only. Our search for and the existence of analytic solutions does not seem to dependent on our particular choice. We leave the very interesting task of calculating $A(S,U)$ explicitly in this model and searching for analytic dS vacua for future work.

In the string theory construction we have to satisfy certain constraints. In particular we want to have a solution with a small string coupling constant ${\rm Re}(S) \gg 1$ and large volume ${\rm Re}(T) \gg 1$ so that string-loop and $\alpha'$ corrections are suppressed and we can trust our supergravity solution. As we mentioned above we can always rescale our parameters and fields such that the (A)dS vacua are at arbitrary values of $S, T$ and $U$ so that this requirement imposes no real constraint on our solution (see \cite{Blaback:2013qza} for a detailed discussion). Further constraints arise from flux quantization and the D3-brane tadpole cancellation condition. In particular the $F_3$ and $H_3$ flux quanta and therefore the $a_I$'s and $b_I$'s in our model are in a certain normalization integers. Since a constant rescaling of the superpotential $W$ maps solutions to solutions, we can rescale $W$ by a very large number to make all the $a_I$'s and $b_I$'s very close to being integers. Then we could drop the decimals and have a correctly quantized solution that is arbitrarily close to our analytic solution. However, the tadpole cancellation condition that ensures that the positive D3-brane charge induced by the $F_3$ and $H_3$ fluxes does not exceed the negative charge from O3-planes in our model, puts a constraint on one particular combination of the $a_I$'s and $b_I$'s. In order to check whether the flux quantization conditions and the tadpole cancellation condition can be satisfied simultaneously, one would have to carefully map out the entire parameter space of dS solutions in this particular model. We refrain from doing so but point out that for example already in our simplest solutions given in subsection \ref{sec:SimpleSolution}, it is easily possible to satisfy either the flux quantization or the tadpole condition.

The above model is a special case of the much larger class of type IIB compactifications on orientifolds of CY$_3$ manifolds or F-theory compactifications on CY$_4$ manifolds. Such compactifications have one axio-dilaton $S$ and potentially hundreds or thousands of complex structure fields $U_I$ as well as a large number of K\"ahler moduli $T_J$. In this case the overall volume modulus has a scalar potential of no-scale type. Similar to our explicit $STU$ model above, it is known that one can obtain no-scale Minkowski solutions in which $S$ and all $U_I$ have positive masses \cite{Giddings:2001yu}. It follows from our general analysis in section \ref{sec:strategy} that even cases with many K\"ahler moduli give rise to stable no-scale Minkowski vacua, if the superpotential allows a sufficient tuning to set $V_{ab}=0$ or at least make it small. It seems therefore likely that by adding non-perturbative corrections from gaugino condensation on D7-branes or from Euclidean D3-branes for all (or many) of the $T_J$, one can find many more stable dS solutions and it would be very interesting to study this further.

A related model that arises in compactifications of type IIA string theory on $SU(3)$-structure manifolds in the presence of non-perturbative corrections was recently analyzed in \cite{Danielsson:2013rza} and numerical dS vacua were found. The K\"ahler and superpotential for this model are
\bea
K &=& -\log (S + \bar S) -3 \log (T + \bar T) -3 \log (U + \bar U) \,,\\
W &=& (a_0 -3 a_1 U + 3 a_2 U^2 - a_3 U^3)  - S(b_0 -3 b_1 U) + T (c_0 -3 c_1 U)+ A e^{-aS} + B e^{-b T}\, .\nn\quad
\eea
We checked that our method easily gives analytic families of dS vacua for this class of models as well.

We also applied our method to another string inspired 4D STU-supergravity model that was first considered in \cite{Shelton:2005cf}.\footnote{Note that \cite{Shelton:2005cf} uses a slightly different notation for the three moduli fields.} It is obtained by applying T-duality arguments to simple flux compactifications of type II string theory and has the following K\"ahler and superpotential
\bea
K &=& -\log (S + \bar S) -3 \log (T + \bar T) -3 \log (U + \bar U) \,,\\
W &=& P(a_I, U)  - S P(b_I, U) + T P(c_I, U)\, ,\nn\quad
\eea
with the $P(k_I, U)$ given in \eqref{Ps}. In this class of models numerically dS vacua were found in \cite{deCarlos:2009fq, Danielsson:2012by} and it is straight-forward to find families of analytic dS solutions using our method. Recently generalizations of this model with 7 moduli were studied in \cite{Blaback:2013ht} and explicit dS vacua were found. It would be interesting to apply our method to these more general models to better understand these dS vacua and their properties.

\section{No-go theorem for dS vacua as continuous deformations of SUSY Minkowski vacua}\label{sec:nogo}

We are interested in various dS vacua, including realistic dS vacua describing the current state of the universe with a very small value of the cosmological constant. Therefore it would be natural to   start by looking at {\it supersymmetric Minkowski vacua, which are relatively easy to find}, see for example \cite{Kallosh:2004yh} for the $F$-term potentials in $\mathcal{N}=1$ supergravity as well as the corresponding Minkowski solutions in that paper. It would seem most natural to slightly deform the K\"ahler or superpotential of a supersymmetric Minkowski vacuum where all scalars are stabilized, such that there are no flat directions. However, we find that this goal is impossible to achieve: we prove that no infinitesimal deformation of the K\"ahler and/or superpotential can change a supersymmetric Minkowski vacuum without flat directions into a non-supersymmetric vacuum.

Assume that we have K\"ahler and superpotential $K$ and $W$ which give rise to a supersymmetric Minkowski minimum at $X^a = X_\star^a$. Then we have $D_a W|_\star=0$ and $W|_\star=0$, and hence also $\partial_a W|_\star = 0$. The Hessian mass matrix at the minimum is
\be
M^2_{\rm   Minkowski} =
\left.\begin{pmatrix} e^{K}|\partial^2W|^2_{a\bar b} & 0\\  0 &  e^{K}|\partial^2W|^2_{\bar a b}\end{pmatrix}\right|_\star \, .
\label{matrixMink}
\ee
In general the mass matrix is positive-semidefinite, there can be flat directions but no tachyons. We assume that $M^2$ is strictly positive-definite, such that all scalars have a positive mass squared in the vacuum. This gives the condition:
\be
\det \partial^2W|_\star \neq 0\,.\label{eq:assumption}
\ee
Now consider a continuous deformation of the K\"ahler and/or superpotential
\be
W (X^a) \rightarrow W (X^a)+  \epsilon \,\delta W(X^a)\, ,\qquad K(X^a,\bar{X}^{\bar{a}}) \rightarrow K(X^a,\bar{X}^{\bar{a}})+  \epsilon \,\delta K(X^a,\bar{X}^{\bar{a}})\,,
\ee
with $\epsilon \ll 1$ a dimensionless parameter. The potential will have an extremum at a position $X_\star' = X_\star + \mathcal{O}(\epsilon)$.
We can write the critical point equation $\partial_a V =0$ as an eigenvalue equation \cite{Marsh:2011aa}:
\be
{\cal M} \cdot v = 2 |W| v\, ,
\ee
with $v$ a unit vector along the supersymmetry breaking direction and ${\cal M}$ a matrix determined by $D^2 W$ as:
\be
v = \frac 1 {\sqrt 2 |F|}\begin{pmatrix} F_a \\ \bar F_{\bar a}\end{pmatrix}
\, ,
\qquad  {\cal M}  = \begin{pmatrix}0 & e^{-i \theta_W} D^2 W\\ e^{i \theta_W}\bar D^2 \bar W &0\end{pmatrix}\,,\qquad \theta_W = \text{Arg}(W)\,.
\ee
We can make a Taylor expansion in the position of the new minimum $X'_\star = X_\star + \mathcal{O}(\epsilon)$ and to lowest order we find
\be
{\cal M}|_\star \cdot v= \mathcal{O}(\epsilon)\, .
\ee
Since $\det {\cal M} _\star= -\det \left.|\partial^2W|^2\right|_\star = \mathcal{O}(\epsilon^0)$ by assumption, see \eqref{eq:assumption}, we cannot satisfy this equation as $\epsilon\to 0$, regardless of the direction of the unit vector $v$ and hence of the direction of supersymmetry breaking.

Thus we have proven that by continuously deforming the K\"ahler and or superpotential, we can only get supersymmetric Minkowski or AdS solutions, but never a dS minimum or any other non-supersymmetric critical point.

One can still hope for small deformations of Minkowski that give stable dS vacua by circumventing the assumptions of the theorem. Either one starts from non-supersymmetric Minkowski, or from a supersymmetric Minkowski vacuum with one or more flat directions in the mass matrix.
Our general class of models described in section \ref{sec:strategy} allows for a range of values for the cosmological constant $\Lambda$. This range includes positive and negative values as well as the Minkowski limit $\Lambda=0$. For all values of $\Lambda$ and $\epsilon$, supersymmetry is broken by a non-zero F-term for the no-scale field $T$. This is thus in clear agreement with the no-go theorem derived above and it can also be explicitly checked in the examples we discuss in section \ref{sec:STUexamples} (cf. in particular \eqref{eq:W_a}). Since our general construction is invariant under arbitrary rescaling of the superpotential, we could rescale the superpotential and all its derivatives by a positive power of $\epsilon$. This would then lead to a solution in which $W$ as well as all the $F_a$ vanish in the limit $\epsilon \rightarrow 0$ and we find a supersymmetric Minkowski solution in this limit. However, due to the rescaling the mass matrix has zero eigenvalues in this limit so that we are again in agreement with the no-go theorem.

\section{F-term uplifting of supersymmetric Minkowski vacua}\label{sec:Polonyi}

In the previous sections we found a method to overcome the no-go theorem in section \ref{sec:nogo} and construct many metastable dS vacua. By changing parameters, one can continuously interpolate between dS and Minkowski space. However,  in many of these models supersymmetry is strongly broken. Meanwhile, many advanced versions of SUSY phenomenology are based on the assumption that the supersymmetry breaking occurs on a relatively low energy scale, with the gravitino mass $m_{3/2} =e^{K/2}|W| \lesssim 10^{3}$ TeV $\sim 10^{-12}$ in Planck mass units. By tuning model parameters, one can suppress the gravitino mass down to $m_{3/2}  \lesssim 10^{-12}$, but this leads in our general construction to a similar suppression of the mass of no-scale modulus. This is similar to the simplest versions of the KKLT scenario, where the mass of the volume modulus is only few times heavier than the gravitino mass, the height of the KKLT barrier is proportional to $m_{3/2}^{2}$, and the vacuum is destabilized during inflation unless $H \lesssim m_{3/2}$ \cite{Kallosh:2004yh}. This places extremely tight constraints on inflationary models \cite{Kallosh:2007wm}. To avoid this problem, one should find a set of dS vacua, where $m_{3/2}$ is very small and the masses of all other string theory moduli are much greater.

In general, one should be able to do it in the context of the scenario developed above. An advantage of this approach is obvious: If it is successful, we can do everything consistently in the versions of supergravity most closely related to string theory. However, one may also look for alternative approaches which may satisfy all the phenomenological requirements outlined above.

Indeed, the methods developed in the previous sections are quite sophisticated, but finding stable supersymmetric Minkowski vacua is very easy, as we will see shortly. We do not live in Minkowski space, but the difference between zero vacuum energy in Minkowski space and the present vacuum energy density is incredibly small,  $V\sim 10^{{-120}}$. And if our main goal is to have dS vacua with tiny vacuum energy and tiny supersymmetry breaking, then it is tempting to consider a more phenomenological approach. One may find strongly stabilized supersymmetric Minkowski vacua first, and then continuously deform the theory to slightly uplift these vacua. One should do it in such a way that the scale of the supersymmetry breaking is small (perhaps smaller than $10^{3}$ TeV), but still many orders of magnitude higher than the minuscule energy scale corresponding to the present vacuum energy density  $V\sim 10^{{-120}}$.

Here we will review a simple implementation of the above procedure in the KL model, which is a generalization of the KKLT scenario  \cite{Kallosh:2004yh,Dudas:2012wi}, and then we will show that a similar mechanism works in more complicated models with many string theory moduli, such as the STU model.

\subsection{KL model and a Polonyi field}
The first example to be considered here is the KL model \cite{Kallosh:2004yh} with $K=  -3 \log (T + \bar T)$ and the racetrack potential
\be
W_{\rm KL}(T) = W_0 + Ae^{-aT}- Be^{-bT} \ .
\label{adssuprace}
\ee
The term $Be^{-bT}$ allows the new model to have a supersymmetric Minkowski solution. Indeed,
for the particular choice of $W_0$,
\be\label{w0}
W_0= -A \left({a\,A\over
b\,B}\right)^{a\over b-a} +B \left ({a\,A\over b\,B}\right) ^{b\over b-a} ,
\ee
the potential of the field $T$ has a supersymmetric minimum $T_{0}= {1\over a-b}\ln \left ({a\,A\over b\,B}\right)$ with
$W_{\rm KL} (T_{0})=0$,  $D_\rho W_{\rm KL}(T_{0}) = 0$,   and $V(T_{0})=0$.

To achieve supersymmetry breaking  one can add to this model the Polonyi field $C$.
The K\"ahler and superpotential are
\be\label{pol}
K = K(T) + C \bar C - \frac{(C \bar C)^{2}}{L^2}\ , \qquad W = W(T) +\mu_{1}+ \mu_2 C\ .
\ee
Here $\mu_{i}$ are supposed to be very small.  Depending on the relation between $\mu_{i}$, this may either lead to a downshift of the Minkowski minimum, making it AdS (for $\mu_{2}^{2}< 3 \mu_{1}^{2}$), or uplift it to a dS minimum (for $\mu_{2}^{2}> 3 \mu_{1}^{2}$). To obtain a slightly uplifted state with the present value of the cosmological constant $\Lambda \sim 10^{{-120}}$, one should have $\mu_{2}^{2}\approx 3 \mu_{1}^{2}$.
In this case one has $m_{{3/2}}^2= {\mu_{1}^{2} \over 8T_{0}^{3}} \ll \mu_{1}^{2} \ll 1$ and $m_{C}^{2} = {12 m_{{3/2}}^2\over L^{2}} \gg m^{2}_{3/2} $ for $L \ll 1$ \cite{Dudas:2012wi}.

Note that this mechanism may lead to the hierarchy of scales in SUSY phenomenology, often associated with split supersymmetry: gluino much lighter than the gravitino, which, in turn, is much lighter than the Polonyi field mass $m_{C}$. This helps to solve many different cosmological problems such as the gravitino problem and the cosmological moduli problem \cite{Dudas:2012wi}.

One may wonder how this mechanism avoids the no-go theorem established earlier. In order to understand it, consider the theory of the fields $T$ and $C$ for $\mu_{i} = 0$. The potential of this theory has a supersymmetric minimum at $T = T_{0}$, with a flat direction $C$. Thus the possibility to uplift this theory to a theory with a stable dS vacuum does not contradict the no-go theorem.

The basic feature of the models of this class is that each of the fields in the model is strongly stabilized. The field $T$ has a very large mass, which does not allow the field $C$ to affect it. The field $C$ is also strongly anchored close to $C =0$ for $L \ll 1$ because the term $\frac{(C \bar C)^{2}}{L^2}$ in the \K\ potential gives it a large mass, and limits the field range to $|C| < L/2\ll 1$ by making the potential singular at $|C| = L/2$.

\subsection{STU model and a Polonyi field}

Our investigation shows that the situation in the STU model is very similar.
One of the simplest examples of the STU model with a Minkowski vacuum with all moduli stabilized is
\be\label{simplemin}
W = A\,(S - S_{0}) (1 - c\, e^{-a\,T}) + B\,(U - U_{0})^2 \ .
\ee
The potential has a supersymmetric minimum at $S = S_{0}$, $U = U_{0}$ and $T = {\log c\over a}$. The masses of all fields in the Minkowski vacuum are given by
\be
m_{s}^{2} = {a^3 A^2\, S_{0}\over 24\, U_{0}^3 \log c}\, , \qquad m_{t}^{2} = {a^{3}\, A^2\, S_{0}\over 24\,   U_{0}^{3} \log c}\, , \qquad  m_{u}^{2} = {a^3 B^2\, U_{0}\over 18\,  S_{0}\,   \log^{3} c}\, .
\ee
Just as in the KL model, one can uplift this stable Minkowski vacuum to a stable dS vacuum by adding the Polonyi field $C$ as we did in (\ref{pol}) with small parameters $\mu_{1} \sim \mu_{2}$, and with $L \ll 1$.

Another simple example is the STU model
\be
W = W_{\rm KL}(T)+ P\,(S - S_{0})^{2} + Q\,(U - U_{0})^2 \ ,
\ee
where $W_{\rm KL}(T)$ is the KL superpotential (\ref{adssuprace}), (\ref{w0}); $P$ and $Q$ are some constants. One can easily check that it has a supersymmetric Minkowski vacuum with all moduli stabilized at $S = S_{0}$, $U = U_{0}$ and $T =  {1\over a-b}\ln \left ({a\,A\over b\,B}\right)$, which can be uplifted to a stable dS vacuum as in the case presented above.

Many other families of supersymmetric Minkowski vacua solutions are possible in the STU model, and in other models with more moduli and fully stabilized supersymmetric Minkowski vacua. An advantage of such models is the possibility to have a very small supersymmetry breaking and a gravitino mass that is unrelated to the strength of the stabilization in the SUSY Minkowski vacuum. This is important for vacuum stabilization during inflation: In the standard version of the KKLT scenario, the Hubble constant during inflation was bounded from above by the gravitino mass \cite{Kallosh:2004yh}. In the KL theory, this constraint disappears, but this requires to introduce an additional exponent  $Be^{-bT}$ in the superpotential. Now we see that in the theory with many moduli, one can achieve this goal in many different ways, even without introducing the second exponent, see (\ref{simplemin}).

There are many other possible ways to achieve uplifting of Minkowski and AdS vacua in supergravity, see for instance
\cite{Achucarro:2006zf}. The main new ingredient that we add to these procedures is finding and uplifting a broad landscape of stable supersymmetric Minkowski vacua. To find such vacua, one should solve a series of equations $W = 0$ and $DW = 0$ with respect to each of the string theory moduli. A large set of such solutions should exist if the superpotential depends on sufficiently many parameters. Among these vacua, there are some that have large positive eigenvalues of the mass matrix. By adding to such models the Polonyi field as shown above, one can uplift these vacua and break supersymmetry without affecting the stability of the original Minkowski landscape.

In this paper we gave only some simple examples of this construction, but it is easy to present more general versions of this construction in the STU model and beyond.

\section{Discussion}\label{sec:discussion}

In this paper we have given a general recipe for constructing abundant families of analytic de Sitter vacua in supergravity models that are motivated by compactifications of string theory. Our method for constructing these de Sitter vacua requires one to impose a hierarchy for the supersymmetry breaking and use the no-scale $T$ modulus in such a way that the approximate no-scale condition effectively removes the negative $-3 |W|^2$ term from the potential
\be
|D_T W|^2 - 3 |W|^2  \sim \mathcal{O}(\epsilon^2)\,,
\label{difT}\ee
with $\epsilon \ll |W|$. The remaining part of the potential for the $X^i$ fields is positive and small
\be
|D_iW|^2  \sim \mathcal{O}(\epsilon^2)\,.
\ee
We evaluate the full mass matrix under the conditions that
\bea
V&=& \Lambda\,  ,\qquad  \partial_a V=0\, , \qquad D_a W =  F_a(\epsilon)\, , \qquad \partial_a \partial_b V = V_{ab}(\Lambda,\epsilon)\, ,
\label{eqs}\eea
where  $F_a(\epsilon)$ is specified in \rf{FT}.  When $V_{ab} = 0$ can be enforced or $V_{ab}$ can be made small, which is possible whenever we can tune sufficiently many third derivatives of $W$, then there exists an abundance of analytic, locally stable families of dS solutions without tachyons. This is based on our careful analysis of the mass matrix in the presence of the two parameters $\epsilon$ and $\Lambda$ as detailed in section \ref{sec:strategy}. Under the above assumption on $V_{ab}$ we find that it is always possible to preserve stability in a region of $\epsilon,\Lambda$ parameter space that is specified in eqs. \rf{cond1}  and  \rf{cond2}. We have used our general method for constructing abundant locally stable (A)dS vacua to obtain explicit examples of families of (A)dS vacua.  These examples confirm our general procedure and the argument that under the conditions \rf{cond1}  and  \rf{cond2} on $\epsilon,\Lambda$ our newly constructed parts of the de Sitter landscape have analytic, locally stable dS solutions, as predicted.

It is important to stress here that we have established the existence of a completely general class of locally stable de Sitter vacua in supergravity  in a region of $\epsilon,\Lambda$ parameters under the conditions that one of the fields has a no-scale \K\, potential and that $W$ is sufficiently generic to allow a tuning of $V_{ab}$. This is to be contrasted with the various earlier attempts to construct de Sitter vacua starting with specific string theory compactifications satisfying the relevant tadpole and quantization conditions, known as  `hunting for de Sitter vacua', see for example \cite{Danielsson:2011au}, \cite{Danielsson:2013rza}. This direct approach did not have a significant rate of success since generically many more de Sitter solutions have been found which have tachyons. For example, in \cite{Danielsson:2013rza} for $10^5$
random choices, there were two stable dS critical points with positive mass matrix and the correct sign of the tadpoles. The new challenge, now that we know how to get locally stable de Sitter vacua in supergravity without tachyons, is to design a new search within string theory models.

For example the relation between our  STU supergravity examples and an explicit string theory compactification deserves further study. However, it is clear that these particular STU-examples are just the simplest ones out of a very large class of string compactifications for which our procedure can be used. It seems plausible that more general string theory examples with a larger number of moduli can accommodate a small cosmological constant more easily and allow us to satisfy all potential string theoretical constraints.  We like to stress that, although we have mostly focused on the two parameters $\Lambda$ and $\epsilon$, we have found analytic solutions that have many more parameters. In general, one expects in addition to $\Lambda$ and $\epsilon$ further parameters that arise from the unit vector $\mu_i$ and that the conditions one has to impose on $W$ do not fix all its parameters. So our method should generically lead to dS solutions with a high dimensional parameter space and a variety of different features. The only universal property that seems to arise in our constructions is that the mass of the no-scale field is always smaller than the gravitino mass. We hope that the main observation of this paper as to why we were able to construct many analytic de Sitter vacua will help to find many more realistic solutions in the future.

We have also discussed an alternative more phenomenological way to produce a large  part of the dS landscape in supergravity where generic locally stable dS vacua originate from supersymmetric stable Minkowski vacua due to the presence of an extra Polonyi superfield. In these models it is easy to describe dS vacua with a gravitino mass that is much smaller than the masses of the scalar fields and that is tunable independently of the value of the cosmological constant and the values of the moduli masses.

All our examples of locally stable dS vacua have a limit to Minkowski space which is either non-supersymmetric or has flat directions.
This fact is in agreement with our general no-go theorem  about the relation between locally stable dS vacua and Minkowski vacua.

\subsubsection*{Acknowledgments}

We acknowledge stimulating discussions with Ulf Danielsson, Shamit Kachru, Liam McAllister, Hiroshi Ooguri, Diederik Roest, Eva Silverstein, Gonzalo Torroba, Thomas Van Riet and Alexander Westphal and particularly thank Ulf Danielsson, Liam McAllister, Diederik Roest, Eva Silverstein and Thomas Van Riet for valuable comments on a draft of this paper. We are supported by the SITP, and by the NSF Grant PHY-1316699 and RK is also supported by the Templeton foundation grant `Quantum Gravity Frontiers'.  BV gratefully acknowledges the Fulbright Commission Belgium and B.A.E.F.\ for support. TW is supported by a Research Fellowship (Grant number WR 166/1-1) of the German Research Foundation (DFG).

\appendix
\section{Conventions}\label{app:conventions}

We need the potential and its first and second derivatives
\begin{eqnarray}\label{appendixpotential}
V=e^{K}\Big(\sum_{a=1}^{n}|D_{a}W|^{2}-3|W|^{2}\Big)\ .
\end{eqnarray}
Following \cite{Denef:2004cf}, the first order derivatives of the potential are:
\begin{eqnarray}\label{DV}
\partial_{a}V=e^{K}\Big((D_{a}D_{b}W)\bar{D}^{b}\bar{W}-2(D_{a}W)\bar{W}\Big)=0\ ,\label{eq:dV_W}
\end{eqnarray}
and the second-order derivatives are:
\begin{eqnarray}
&&D_{a}\partial_{b}V=e^{K}\Big((D_{a}D_{b}D_{c}W)\bar{D}^{c}\bar{W}-(D_{a}D_{b}W)\bar{W}\Big)\ ,\label{DDV1}\\
&&{D}_{a}\bar\partial_{\bar b}V=e^{K}\Big(-{R}_{a\bar{b}c\bar{d}}(\bar{D}^{c}\bar{W}) D^{\bar{d}}W+G_{a \bar b}(D_{c}W)
\bar{D}^{c}\bar{W}-({D}_{a}{W})\bar D_{\bar b} \bar W\nonumber\\
&&\hspace{2.5cm}+(D_{a}D_{c}W)(\bar{D}_{\bar{b}}\bar{D}^{c}\bar{W})-2G_{a\bar{b}}|W|^{2}\Big)\,,\label{DDV2}
\end{eqnarray}
where the derivative $D$ is K\"ahler  covariant and covariant on the scalar manifold, for example $D_a W = \partial_a W + K_a W$. For an extremum of the potential we have $d V = 0$ and hence $Dd V = d^2 V$.

Here and in the following, we  denote partial derivatives by subscripts: $K_{ab} = \partial_a \partial_b K,W_{a} = \partial_a W$ and so on. Indices are raised with the inverse scalar metric, as in the following definition
\begin{equation}
\bar D^a = G^{a \bar b}\bar D_{\bar b}\, .
\end{equation}
The metric and curvature of the K\"ahler geometry are
\begin{align}
 G_{a \bar b} &= K_{a \bar b}\ , \qquad \Gamma_{ab}^{c} = G^{c\bar d } K_{ab\bar d}\ ,\\
 R_{a\bar b c \bar d} &= K_{a\bar b c \bar d} - K_{\bar b \bar d  e} G^{e \bar f} K_{a c \bar f}\ .
\end{align}

\section{Positive-definiteness of diagonal blocks of mass matrix}\label{sec:proof}

In this section we study the necessary and sufficient conditions for the holomorphic-anti-holomorphic part of the mass matrix to be positive-definite, $V_{a\bar{b}}>0$. Note that this condition $V_{a\bar{b}}>0$ is only sufficient for stability of the entire mass matrix, if the entries of $V_{ab}$ are sufficiently small.

The fields are split as $X^a = (T,X^i)$, with a \K\ potential $K = -3 \log(T+\bar T) + K(X^i,\bar{X}^{\bar{i}})$, where $K(X^i,\bar{X}^{\bar{i}})$ is completely arbitrary and can depend on any number of fields. We will make extensive use of the geometric properties, such as the no-scale property and the Riemann curvature:
\be\label{eq:Riemann}
K_T K^T = 3\ , \qquad R_T{}^T{}_T{}^T = \frac 23\ , \qquad R_{T \bar{T} i \bar{j}}=0\ .
\ee
We show now that for a particular ansatz for $D_a W  = F_a$ the matrix $V_{a  \bar b}$ is always positive-definite, if the norm $|F_i|=\sqrt{\bar F^i F_i}$ is  small enough. We introduce a parameter $\epsilon = |F_i|$ and write:
\be
F_T = F_T(\epsilon)\,, \qquad F_i = \epsilon \mu_i\, ,
\label{eq:ansatz}
\ee
with $\bar \mu^i$  a unit vector $|\mu_i|^2 = \bar \mu^i \mu_i = 1$.
The component $F_T$ is not independent, as it is related to $F_i$ and the cosmological constant through the condition $V=\Lambda$ of \eqref{appendixpotential}.

For $n$ fields $X^a$, $V_{a \bar b}$ is $(n\times n)$-dimensional and hard to analyze. We will rewrite it in terms of a  two-by-two matrix.
The matrix $V_{a \bar b}$ is positive-definite provided $V_{a \bar b}  \bar v ^av^{\bar b}>0$ for all vectors $v_a$. We can always write an arbitrary vector as $v_a = \alpha L_a + \beta F_a$ for some complex  numbers $\alpha,\beta$ and $L_a$ a unit vector  orthogonal to the supersymmetry breaking direction:
\be
L_a   \bar F^a = 0\ , \qquad |L_a|^2 = \bar L^a L_a= 1\ .
\ee
This gives the useful relation:
\be
L_T = -\epsilon \frac{L_i \bar \mu^i}{\bar F^T }\ .\label{eq:GF}
\ee
Then the condition $V_{a \bar b}  \bar v ^av^{\bar b}>0$ is equivalent to the matrix $P(L_a)$ being positive-definite for any choice of vector $L_a$, with
\be
P(L_a) = \begin{pmatrix}\bar F^a F^{\bar b} V_{a \bar b}& \bar L^a F^{\bar b} V_{a \bar b}\\
L^{\bar a} \bar F^{ b} V_{\bar a b}&\bar L^a L^{\bar b} V_{a \bar b}\end{pmatrix}\ .
\ee
A matrix is positive-definite if all determinants of its upper-left square sub-matrices are positive, which gives the conditions:
\begin{align}
0<\ &P_{1\bar 1}= \bar F^a F^{\bar b} V_{a \bar b} \ , \label{upper}\\
0<\ & \det P = (\bar F^a F^{\bar b} V_{a \bar b})(\bar L^a L^{\bar b} V_{a \bar b}) - |\bar L^a F^{\bar b} V_{a \bar b}|^2\ .\label{eq:cond-pos}
\end{align}
Note that this trivially implies:
\be
0< P_{2\bar 2}= \bar L^a L^{\bar b} V_{a \bar b} \ .\label{lower}\\
\ee
We investigate the conditions \eqref{upper}, \eqref{eq:cond-pos}, \eqref{lower} for $F_a$ obeying the ansatz \eqref{eq:ansatz}. Using the expression for $V_{a\bar b}$ in terms of the superpotential \eqref{DDV2}, the components of  $P(L_a)$ are:
\begin{align}
 \bar F^a F^{\bar b} V_{a \bar b} =\  & e^{K}\Big(-{R}_{a\bar{b}c \bar d}\bar F^aF^{\bar b}  \bar F^c F^{\bar d} + 2 |F_a|^2|W|^2\Big)\ ,\label{eq:GoldstrinoMass}\\
\bar L^a L^{\bar b} V_{a \bar b} =\ & e^{K}\Big(-{R}_{a \bar b c \bar{d}}  \bar L^{a} L^{\bar b }\bar  F^cF^{\bar d} +|L_a|^2 (e^{-K}\Lambda +  |W|^2)+|L\cdot D^2W|^2\Big{)}\ ,\nonumber\\
 \bar L^a F^{\bar b} V_{a \bar b} =\ & -e^{K}{R}_{a\bar{b}c \bar d} \bar L^aF^{\bar b}\bar F^c F^{\bar d} \ .\nonumber
\end{align}
To obtain  these expressions, we used $L_a \bar F^a =0$, the condition $\partial_a V =0$ as in \eqref{eq:dV_W} and $V=\Lambda$.

The sgoldstino component, the combination $\bar F^a F^{\bar b} V_{a \bar b}$, can be expressed through $\Lambda = e^K(|F_a|^2  - 3|W|^2) = e^K(|F_T|^2 + \epsilon^2 |\mu_i|^2 - 3|W|^2)$ and \eqref{eq:Riemann}, \eqref{eq:ansatz} as
\bea
e^{-K} \bar F^a F^{\bar b} V_{a \bar b} &=& 2 |W|^2(e^{-K} \Lambda + 3 |W|^2) -\frac23 (e^{-K} \Lambda + 3 |W|^2 - \epsilon^2 |\mu_i|^2)^2 -\epsilon^4\bar \mu^i \mu^{\bar j} \bar \mu^k \mu^{\bar \ell} R_{i\bar j k \bar \ell} \cr
&=&2 |W|^2 (2 \epsilon^2 |\mu_i|^2 -e^{-K} \Lambda)- \frac 23 (e^{-K} \Lambda - \epsilon^2 |\mu_i|^2)^2  -\epsilon^4\bar \mu^i \mu^{\bar j} \bar \mu^k \mu^{\bar \ell} R_{i\bar j k \bar \ell}\,.
\label{FFV}
\eea
Only the very first term in this equation is always positive. By taking it larger than the last term, we can have positivity of \rf{FFV} for \emph{any choice} of  the $X^i$ scalar geometry. Recalling that $|\mu_i|^2=1$ this requires
\be
\epsilon\ll |W|\,.
\ee
Then we still need that the terms involving the cosmological constant do not give too large negative contributions. In the approximation $\epsilon\ll |W|$ the curvature and the other $\epsilon^4$ term can be dropped, leaving a quadratic inequality for $\Lambda$. Dropping subleading terms in the $\epsilon/ |W|$ expansion, we find the upper bound
\be\label{eq:lambda}
\lambda \equiv \frac{e^{-K} \Lambda}{2\epsilon^2 |\mu_i|^2} <1\  .
\ee
We next discuss the positivity condition  \eqref{lower} $L^a L^{\bar b} V_{a \bar b}$. Only the curvature term can be negative. By the choice of K\"ahler potential \eqref{eq:SeparableKahler} all mixed $T,X^i$ components of the Riemann tensor are zero (cf. \eqref{eq:Riemann}). Using \eqref{eq:GF} we then find that the curvature term is subleading in the $\epsilon/ |W|$ expansion
\begin{align}\label{first}
e^{-K} \bar L^a L^{\bar b} V_{a \bar b}
=\ & |W|^2 |L_a|^2+|L\cdot D^2W|^2+2\lambda \epsilon^2 |\mu_i|^2|L_a|^2-\frac 23  \epsilon^2 (\bar \mu^i L_i)^2- \epsilon^2{R}_{   i\bar  j k \bar{\ell}}  \bar L^{i} L^{\bar j}\bar  \mu^k\mu^{\bar \ell}\,,
\end{align}
and we find that this is positive for small enough $\epsilon/|W|$.

In a similar way we can write $e^{-K}\bar L^a F^{\bar b} V_{a \bar b} $ as
\begin{align}
e^{-K}\bar L^a F^{\bar b} V_{a \bar b} =\ & 2 \epsilon \bar L^i \mu_i|W|^2  -  \frac 23 \epsilon^3 \bar L^i \mu_i |\mu_i|^2 (1 - 2\lambda ) -  \epsilon^3 \bar L^{i} \mu^{\bar j} \bar \mu^k  \mu^{\bar \ell}  R_{i \bar j k \bar \ell}  \ .
\end{align}
In short, we find for $\epsilon \ll  |W|$ that
\be
P(L_a) = e^K |W|^2\begin{pmatrix}  4\epsilon^2( 1-\lambda)|\mu_i|^2 & 2 \epsilon \bar L^i \mu_i\\2 \epsilon  L_i \bar\mu^i & |L_a|^2+\frac{|L\cdot D^2W|^2}{|W|^2}\end{pmatrix} + \mathcal{O}(\epsilon^4)\ .
\ee
This gives for the  determinant of $P(L_a)$:
\be
\det P(L_a)= 4\epsilon^2 e^{2K} |W|^4\left[ (1 - \lambda)|\mu_i|^2\left(|L_a|^2 +  \frac{|L\cdot D^2W|^2}{|W|^2} \right)-|\bar \mu^i L_i|^2\right]+\mathcal{O}(\epsilon^4)\  .
\label{det}\ee
The term $-|\bar \mu^i L_i|^2$ gives a negative contribution and hence $\det P(L_a)>0$ is a stronger condition than $\lambda <1$ as expected.

We can now summarize all conditions on $\Lambda$ and $\epsilon$ for the positivity of $V_{a\bar b} >0$, irrespective of the K\"ahler geometry of the scalars $X^i$. First we require
\be
\epsilon \ll |W|\ .\label{eq:epscond}
\ee
We need for consistency that $\Lambda = 2 \lambda e^{K}\epsilon^2 |\mu_i|^2$ and $\lambda$ is bounded by $\lambda <1$, which is subsumed by
$\det P >0$  \eqref{det} in the superpotential-dependent condition
\be
\lambda < \min_{\{L_a\}} \left(1 - \frac{|\bar \mu^i L_i|^2/|\mu_i|^2}{|L_a|^2 +  \frac{|\bar L^a D_a D_bW|^2}{|W|^2}}\right) = \min_{\{L_a\}} \left(1 - \frac{|\bar \mu^i L_i|^2}{1 +  \frac{|\bar L^a D_a D_bW|^2}{|W|^2}}\right)\\ , \label{eq:lambdacond}
\ee
where we made a simplification by choosing $\{ L_a \}$ a set of unit vectors that are all orthogonal to $F_a$. Because $\mu_i$ and the $L_a = \{-\epsilon L_i \bar \mu^i/\bar F^T, L_i\}$ are unit vectors, we find that $|\bar \mu^i L_i|<1$ as long as $\epsilon \neq 0$ so that there is always a range of small positive $\lambda$ for which $V_{a\bar b}>0$.

Finally, note that \eqref{eq:lambda} together with $\Lambda = V = e^K(|F_T|^2 + \epsilon^2 |\mu_i|^2- 3 |W|^2)$ and $\epsilon \ll |W|$ implies that
\be
F_T = K_T W + \nu_T \frac{\epsilon^2}{|W|^2}\ ,
\ee
with $\nu_T = \mathcal{O}[(\epsilon/|W|)^0]$.\footnote{Note that the complex $\nu_T$ has to satisfy the real constraint $\Lambda = V = e^K(|F_T|^2 + |F_i|^2 - 3 |W|^2)$, giving $\Lambda= e^K(2 {\rm Re} (\nu_T \epsilon^2/W) + \nu_T^ 4 \epsilon^2|W|^4 + |F_i|^2$).}
Hence the requirement of positive-definite $V_{a \bar b}$ regardless of the $X^i$ geometry implies that the supersymmetry breaking direction $F_a$ is dominantly in the no-scale direction since $|F_T| \sim |W| \gg \epsilon \sim |F_i|$.

\section{Details of the analytic dS solutions in Class II}\label{app:solutions}

Here we present a solution for the model discussed in subsection \ref{sec:SimpleSolution} where all $a_I, b_I,c_I$ with $I=0,1,2$ and $a_3$ are given in terms of $a, b_3, c_3, \epsilon, \lambda$. The coefficient $b_3, c_3$ are unconstrained and we could set them to zero $b_3= c_3=0$.
\begin{align}
 a_0&=a_3 + \frac{6 \e^2 (3+\e^2+2\l)-a(2-\e^2(10-3\l)-\l(7-6\l))}{24 a}\,,\\
 a_1&=a_3+\frac{(1+\e)(2\e(1+\e+\e^2)-a(1-\e(2+\e)))+\l(a(3-\e)(1+\e)+4\e(2+\e))-2a \l^2}{12a}\,, \\
 a_2&=\frac13(-a_0+3a_1+a_3)+\frac{a+\e^2(6+a)-2a\l}{18a}\,,\\
 a_3&=b_3-e^{-a} c_3 -\frac{a^2 (1-\e^2-2 \l)^3 - a (1-\e^2-2 \l)^2 (\e (3+5 \e)+6 \l)+2\e(1+\e)^2(1+3\e(1+\e(1-\e)))}{48\e}\nn\\
 &\quad+\frac{\l(1-\e(2+\e(6-\e(6+7\e)))-4\l+2\e \l (3+5\e)+4\l^2)}{12\e}\,,\\
 b_0&=b_3 -\frac{1+\e^2(4+9\l)+3\e^4-\l(5-6\l)}{24}\,,\\
 b_1&=b_3-\frac{\e(1+\e)(1+\e^2)-\l(1-\e(2+3\e))+2\l^2}{12}\,,\\
 b_2&=\frac13(-b_0+3b_1+b_3)-\frac{1+\e^2-2\l}{18}\,,\\
 c_0&=c_3 -\frac{e^a (a^2(1-\e^2-2\l)^2+2\e^2(3+\e^2+2\l)-a(1-\e^2-2\l)(1+3\e^2+4\l)}{8a}\,,\\
 c_1&=\frac13(2c_0+c_3)+\frac{e^a \e(a(1-\e^2-2\l)-1+\e(1-2\e)-4\l)}{6a} \,,\\
 c_2&=\frac13(-c_0+3c_1+c_3) -\frac{e^a \e^2}{3a}\,.
\end{align}

\end{document}